\documentclass{Pos}

\title{Theoretical aspects of electroweak symmetry breaking in SUSY models}

\ShortTitle{Theoretical aspects of electroweak symmetry breaking in SUSY models}

\author{\speaker{Roman Nevzorov}%
         \thanks{On leave of absence from the Theory Department, ITEP, Moscow, Russia.}\\
        University of Hawaii\\
        E-mail: \email{nevzorov@phys.hawaii.edu}}

\abstract{The electroweak (EW) symmetry breaking in the simplest supersymmetric (SUSY)
extensions of the standard model (SM), i.e. minimal and next-to-minimal supersymmetric
standard models (MSSM and NMSSM), is considered. The spectrum of the Higgs particles,
upper bound on the mass of the lightest Higgs boson and little hierarchy problem
are discussed. The breakdown of gauge symmetry and Higgs phenomenology within
the $E_6$ inspired SUSY models with extra $U(1)'$ factor are briefly reviewed.}

\FullConference{The XIXth International Workshop on High Energy Physics and Quantum Field Theory \\
                 8-15 September 2010\\
                 Golitsyno, Moscow, Russia}

\begin{document}

\section{Introduction}
Nowadays, there are serious reasons to believe that new particles and interactions
should exist at the TeV scale. We expect that LHC might be able to shed light on
the physics beyond the Standard Model (SM), origin of dark matter and the mechanism
of electroweak (EW) symmetry breaking in the near future. At the moment the
supersymmetric (SUSY) models are the best motivated extensions of the SM.
SUSY models are defined by the field content, structure of gauge interactions and
superpotential. In order to give both up- and down-type fermions a mass and to ensure 
anomaly cancellation the Higgs sector of the minimal supersymmetric standard 
model (MSSM) includes two Higgs doublets ($H_1$ and $H_2$) with the opposite hypercharge.
Also in SUSY models R-parity is normally imposed $\left(R=(-1)^{3(B-L)+2S}\right)$.
It forbids the baryon and lepton number violating operators that lead to rapid proton
decay. R-parity ensures that the lightest supersymmetric particle (LSP) is absolutely 
stable and can play the role of dark matter. The most general renormalizable superpotential 
of the MSSM can be presented in the following form
\begin{equation}
W_{MSSM} = \epsilon_{ij}(y^U_{ab}Q_a^j u^c_bH_2^i +
y^D_{ab}Q_a^j d^c_b H_1^i + y^L_{ab} L_a^j e^c_b H_1^i + \mu H_1^iH_2^j),
\label{3}
\end{equation}
where $i,j=1,2,3$ are the $SU(2)$ and $a,b=1,2,3$ are the generation indices.
In Eq.~(\ref{3}) colour indices are suppressed.

Since SUSY particles are considerably heavier than observed quarks and
leptons supersymmetry must be broken. Thus the Lagrangian of SUSY models
can be written as
\begin{equation}
{\cal L}={\cal L}_{SUSY}+{\cal L}_{soft}\,.
\label{5}
\end{equation}
where ${\cal L}_{soft}$ contains soft SUSY breaking terms that break supersymmetry but
do not induce quadratic divergences that get cancelled automatically within the models
based on the exact global supersymmetry. In the MSSM the set of the soft SUSY breaking
terms includes
\begin{equation}
\begin{array}{lcr}
-{\cal L}_{soft}& = &\sum_{i}^{}m^2_i |\varphi_i|^2+(\frac{1}{2}
\sum_{\alpha}^{} M_{\alpha}\tilde{\lambda}_{\alpha}\tilde{\lambda}_{\alpha}
+ \sum_{a, b} [A^U_{ab} y^U_{ab}\tilde{Q}_a\tilde{u}^c_b H_2\\[2mm]
&+& A^D_{ab} y^D_{ab}\tilde{Q}_a\tilde{d}^c_b H_1+
A^L_{ab} y^L_{ab} \tilde{L}_a\tilde{e}^c_b H_1]
+ B\mu H_1 H_2 +h.c.)\,,\\
\end{array}
\label{6}
\end{equation}
where $\varphi_i$ are scalar components of chiral superfields
while $\tilde{\lambda}_{\alpha}$ are gaugino fields.
To avoid fine--tuning SUSY breaking mass parameters are expected to be
in the TeV range.

\section{EW symmetry breaking in the MSSM}
Including soft SUSY breaking terms and radiative corrections,
the Higgs effective potential in the MSSM can be written as
\begin{eqnarray}
V=m_1^2|H_1|^2+m_2^2|H_2|^2-m_3^2(H_1 H_2+h.c.)+ \frac{g_2^2}{8}\left(H_1^+\sigma_a H_1+H_2^+\sigma_a
H_2\right)^2\nonumber\\
+\frac{{g'}^2}{8}\left(|H_1|^2-|H_2|^2\right)^2+\Delta V\,,\qquad\qquad\qquad\qquad\qquad
\label{7}
\end{eqnarray}
where $g'=\sqrt{3/5} g_1$, $g_2$ and $g_1$ are the low energy (GUT normalised)
$SU(2)_W$ and $U(1)_Y$ gauge couplings, $m_1^2=m_{H_1}^2+\mu^2$, $m_2^2=m_{H_2}^2+\mu^2$
and $m_3^2=-B\mu$. In Eq.~(\ref{7}) $\Delta V$ represents the contribution of loop
corrections to the Higgs effective potential.

At the physical minimum of the scalar potential (\ref{7}) the Higgs
fields develop vacuum expectation values (VEVs)
\begin{equation}
<H_1>=\frac{1}{\sqrt{2}}\left(
\begin{array}{c}
v_1\\ 0
\end{array}
\right) , \qquad
<H_2>=\frac{1}{\sqrt{2}}\left(
\begin{array}{c}
0\\ v_2
\end{array}
\right)\,.
\label{8}
\end{equation}
breaking the $SU(2)_W\times U(1)_Y$ gauge symmetry to $U(1)_{em}$ associated
with electromagnetism and generating the masses of all bosons and fermions.
The value of $v=\sqrt{v_1^2+v_2^2}\simeq 246\,\mbox{GeV}$ is fixed
by the Fermi scale. At the same time the ratio of the Higgs VEVs remains
arbitrary. Hence it is convenient to introduce $\tan\beta=v_2/v_1$.

The vacuum configuration (\ref{8}) is not the most general one. Because of
the $SU(2)$ invariance of the Higgs potential (\ref{7}) one can always make
$<H_2^{+}>=0$ by virtue of a suitable gauge rotation. Then the requirement
$<H_1^{-}>=0$, which is a necessary condition to preserve $U(1)_{em}$ in the 
physical vacuum, is equivalent to requiring the squared mass of the physical 
charged scalar to be positive. It imposes additional constraints on 
the parameter space.

At tree--level $\Delta V=0$ and the Higgs potential is described by the sum of
the first five terms in Eq.~(\ref{7}). Notice that the Higgs self--interaction
couplings in Eq.~(\ref{7}) are fixed and determined by the gauge couplings in
contrast with the SM. At the tree--level the MSSM Higgs potential contains
only three independent parameters: $m_1^2$,\,$m_2^2$,\,$m_3^2$.
The stable vacuum of the scalar potential (\ref{7}) exists only if
\begin{equation}
m_1^2+m_2^2> 2|m_3|^2\,.
\label{9}
\end{equation}
This can be easily understood if one notice that in the limit $v_1^2=v_2^2$ the
quartic terms in the Higgs potential vanish. In the considered case the scalar
potential (\ref{7}) remains positive definite only if the condition (\ref{9})
is satisfied. Otherwise physical vacuum becomes unstable. On the other hand
Higgs doublets acquire non-zero VEVs only when
\begin{equation}
m_1^2 m_2^2 < |m_3|^4\,.
\label{10}
\end{equation}
Indeed, if $m_1^2 m_2^2 > |m_3|^4$ then all Higgs fields have positive masses
for $v_1=v_2=0$ and the breakdown of EW symmetry does not take place.
The conditions (\ref{9}) and (\ref{10}) also follow from the equations for the
extrema of the Higgs boson potential. At tree--level the minimization conditions
in the directions (\ref{8}) in field space read:
\begin{eqnarray}
\frac{\partial V}{\partial v_1}=\left(m_1^2+\frac{\bar{g}^2}{8}(v_1^2-v_2^2)\right)v_1-m_3^2 v_2=0\,,\nonumber\\
\frac{\partial V}{\partial v_2}=\left(m_2^2+\frac{\bar{g}^2}{8}(v_2^2-v_1^2)\right)v_2-m_3^2 v_1=0\,,
\label{11}
\end{eqnarray}
where $\bar{g}=\sqrt{g_2^2+g'^2}$. Solution of the minimization conditions
(\ref{11}) can be written in the following form
\begin{eqnarray}
\sin 2\beta=\frac{2 m_3^2}{m_1^2+m_2^2}\,,\qquad\qquad
\frac{\bar{g}^2}{4}v^2=\frac{2(m_1^2-m_2^2\tan^2\beta)}{\tan^2\beta-1}\,.
\label{12}
\end{eqnarray}
Requiring that $v^2>0$ and $|\sin 2\beta|<1$ one can reproduce conditions (\ref{9})
and (\ref{10}).

From Eqs.~(\ref{9})--(\ref{12}) it is easy to see that the appropriate breakdown of
the EW symmetry can not be always achieved. In particular the breakdown of
$SU(2)_W\times U(1)_Y \to U(1)_{em}$ does not take place when $m_2^2=m_1^2$.
This is exactly what happens in the constrained version of the MSSM (cMSSM) which
implies that $m_i^2(M_X)=m_0^2$,\,\,$A^k_{ab}(M_X)=A$,\,\,$M_{\alpha}(M_X)=M_{1/2}$.
Thus in the cMSSM $m_2^2(M_X)=m_1^2(M_X)$. Nevertheless the correct pattern
of the EW symmetry breaking can be achieved within this SUSY model.
Since the top--quark Yukawa coupling is large it affects the renormalisation
group (RG) flow of $m_2^2(Q)$ rather strongly resulting in small or even negative
values of $m_2^2(Q)$ at low energies that triggers the breakdown of the
EW symmetry. This is the so-called radiative mechanism of the EW symmetry
breaking (EWSB).

The radiative EWSB demonstrates the importance of loop effects in the considered
process. In addition to the RG flow of all couplings one has to take into account
loop corrections to the Higgs effective potential which are associated with the
last term $\Delta V$ in Eq.~(\ref{7}). In the simplest SUSY extensions of the SM
the dominant contribution to $\Delta V$ comes from the loops involving the top--quark
and its superpartners because of their large Yukawa coupling $h_t$.  In the one--loop 
approximation the contribution of the top--quark and its superpartners to $\Delta V$ 
is determined by the masses of the corresponding bosonic and fermionic states, i.e.
\begin{eqnarray}
\Delta V=\frac{3}{32\pi^2}\left[m_{\tilde{t}_1}^4\left(
\ln\frac{m_{\tilde{t}_1}^2}{Q^2}-\frac{3}{2}\right)+
m_{\tilde{t}_2}^4\left(\ln\frac{m_{\tilde{t}_2}^2}{Q^2}-\frac{3}{2}\right)
-2m_t^4\left(\ln\frac{m_t^2}{Q^2}-\frac{3}{2}\right)\right]\,,\\[2mm]
m_{\tilde{t}_{1,2}}=\frac{1}{2}\left(m^2_Q+m^2_U+2m_t^2 \pm
\sqrt{(m_Q^2-m_U^2)^2+4m_t^2 \, X_t^2}\right)\,,\qquad\qquad\nonumber
\label{14}
\end{eqnarray}
where $X_t=A_t-\mu/\tan\beta$ is a stop mixing parameter, $A_t$ is a trilinear
scalar coupling associated with the top quark Yukawa coupling and $m_t$ is the
running top quark mass
$$
m_t(M_t)=\frac{h_t(M_t)}{\sqrt{2}}v\sin\beta\,.
$$

Initially the sector of EWSB involves eight degrees of freedom. However three
of them are massless Goldstone modes which are swallowed by the $W^{\pm}$ and
$Z$ gauge bosons. The $W^{\pm}$ and $Z$ bosons gain masses via the interaction
with the neutral components of the Higgs doublets so that
$$
M_W=\frac{g_2}{2}v\,,\qquad\qquad M_Z=\frac{\bar{g}}{2}v\,.
$$

When CP in the MSSM Higgs sector is conserved the remaining five
physical degrees of freedom form two charged, one CP--odd and two CP-even
Higgs states. The masses of the charged and CP-odd Higgs bosons are
\begin{equation}
m_{A}^2=m_1^2+m_2^2+\Delta_A\,,\qquad\qquad
M_{H^{\pm}}^2=m_A^2+M_W^2+\Delta_{\pm}\,,
\label{15}
\end{equation}
where $\Delta_{\pm}$ and $\Delta_A$ are the loop corrections.
The CP--even states are mixed and form a $2\times 2$ mass matrix.
It is convenient to introduce a new field space basis $(h,\,H)$
rotated by the angle $\beta$ with respect to the initial one:
\begin{eqnarray}
\mbox{Re} \, H_1^0= (h \cos\beta- H \sin\beta+v_1)/\sqrt{2}\,,\nonumber\\
\mbox{Re} \, H_2^0= (h \sin\beta+ H \cos\beta+v_2)/\sqrt{2}\,.
\label{16}
\end{eqnarray}
In this new basis the mass matrix of the Higgs scalars takes the form
\cite{Kovalenko:1998dc}
\begin{equation}
M^2=\left(
\begin{array}{ll}
M^2_{11}\quad & M^2_{12}\\[3mm]
M^2_{21}\quad & M^2_{22}
\end{array}
\right)=
\left(
\begin{array}{ll}
\displaystyle\frac{\partial^2 V}{\partial v^2} &
\displaystyle\frac{1}{v}\frac{\partial^2V}{\partial v \partial\beta}\\[3mm]
\displaystyle\frac{1}{v}\frac{\partial^2V}{\partial v \partial\beta} &
\displaystyle\frac{1}{v^2}\frac{\partial^2V}{\partial\beta^2}
\end{array}
\right)\,,
\label{17}
\end{equation}
\begin{eqnarray}
M_{22}^2=m_A^2+M_Z^2\sin^2 2\beta+\Delta_{22}\, ,\qquad\quad
M_{12}^2=M_{21}^2=-\frac{1}{2} M_Z^2\sin 4\beta+\Delta_{12}\,,
\label{18}
\end{eqnarray}
$$
M_{11}^2=M_Z^2\cos^2 2\beta+\Delta_{11}\,,
$$
where $\Delta_{ij}$ represent the contributions from loop corrections.
In Eqs.~(\ref{18}) the equations for the extrema of the Higgs
boson effective potential are used to eliminate $m_1^2$ and $m_2^2$.

From Eqs.~(\ref{15}) and (\ref{18}) one can see that at tree--level
the masses and couplings of the Higgs bosons in the MSSM can be
parametrised in terms $m_A$ and $\tan\beta$ only. The masses of the
two CP--even eigenstates obtained by diagonalizing the matrix
(\ref{17})--(\ref{18}) are given by
\begin{equation}
m_{h_1,\,h_2}^2=\frac{1}{2}\left(M^2_{11}+
M^2_{22}\mp\sqrt{(M_{22}^2-M_{11}^2)^2+4M^4_{12}}\right)\,.
\label{19}
\end{equation}
The qualitative pattern of the Higgs spectrum depends very strongly on
the mass $m_A$ of the pseudoscalar Higgs boson. With increasing $m_A$
the masses of all the Higgs particles grow. At very large values of
$m_A$ ($m_A^2\gg v^2$), the lightest Higgs boson mass approaches its
theoretical upper limit $\sqrt{M_{11}^2}$. Thus the top--left entry
of the CP--even mass matrix (\ref{17})--(\ref{18}) represents an upper
bound on the lightest Higgs boson mass--squared. In the leading two--loop
approximation the mass of the lightest Higgs boson in the MSSM does
not exceed $130-135\,\mbox{GeV}$. This is one of the most important
predictions of the minimal SUSY model that can be tested at the LHC
in the near future.

It is important to study the spectrum of the Higgs bosons together
with their couplings to the gauge bosons because LEP set stringent
limits on the masses and couplings of the Higgs states. In the rotated
field basis $(h, H)$ the trilinear part of the Lagrangian, which determines
the interactions of the neutral Higgs states with the $Z$--boson,
is simplified:
\begin{equation}
L_{AZH}=\frac{\bar{g}}{2}
M_{Z}Z_{\mu}Z_{1\mu}h+\frac{\bar{g}}{2}Z_{\mu}
\biggl[H(\partial_{\mu}A)-(\partial_{\mu}H)A\biggr]~.
\label{20}
\end{equation}
In this basis the SM-like CP--even component $h$ couples to a pair of
$Z$ bosons, while the other one $H$ interacts with the pseudoscalar $A$
and $Z$. The coupling of $h$ to the $Z$ pair is exactly the same as in
the SM. The couplings of the Higgs mass eigenstates to a $Z$ pair
($g_{ZZh_i}$, $i=1,2,3$) and to the Higgs pseudoscalar and $Z$ boson
($g_{ZAh_i}$) appear because of the mixing between $h$ and $H$.

Following the traditional notations, one can define the normalised
$R$--couplings as: $g_{VVh_i}=R_{VVh_i}\times \mbox{SM coupling}$
$\left(V=Z,\,W^{\pm}\right)$; $g_{ZAh_i}=\frac{\bar{g}}{2}R_{ZAh_i}$.
The absolute values of all these $R$--couplings vary from zero to unity.
The relative couplings $R_{ZZh_i}$ and $R_{ZAh_i}$ are given in terms
of the angles $\alpha$ and $\beta$:
\begin{equation}
R_{VVh_1}=-R_{ZAh_2}=\sin(\beta-\alpha)\,,\qquad\qquad R_{VVh_2}=R_{ZAh_1}=\cos(\beta-\alpha)\,,
\label{21}
\end{equation}
where the angle $\alpha$ is defined as follows:
\begin{equation}
\begin{array}{rcl}
h_1&=&-(\sqrt{2}\, \mbox{Re}\,H_1^0-v_1)\sin\alpha+(\sqrt{2}\, \mbox{Re}\,H_2^0-v_2)\cos\alpha\,,\\[1mm]
h_2&=&(\sqrt{2}\, \mbox{Re}\,H_1^0-v_1)\cos\alpha+(\sqrt{2}\, \mbox{Re}\,H_2^0-v_2)\sin\alpha\,.
\end{array}
\label{22}
\end{equation}
From Eqs.~(\ref{21}) it becomes clear that in the MSSM the couplings of the
lightest Higgs boson to the Z pair can be substantially smaller than in the SM.
Therefore the experimental lower bound on the lightest Higgs mass in the MSSM is
weaker than in the SM. On the other hand, when $m_A^2\gg v^2$ the lightest
CP--even Higgs state is predominantly the SM-like superposition of the neutral
components of Higgs doublets $h$ so that $R_{ZZh_1}\simeq 1$. In this case
the lightest Higgs scalar has to satisfy LEP constraint on the mass of the SM--like
Higgs boson, i.e. it should be heavier than $114.4\,\mbox{GeV}$.

\begin{figure}
\centering
~\hspace*{-6.1cm}{$m_{h_1}$}\hspace{6.8cm}{$m_{h_1}$}\\
\includegraphics[width=0.43\textwidth]{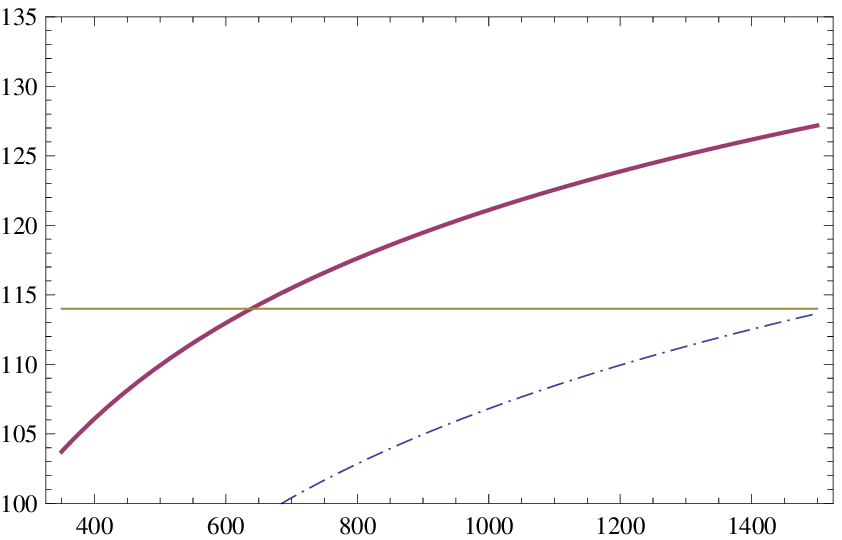}\qquad
\includegraphics[width=0.43\textwidth]{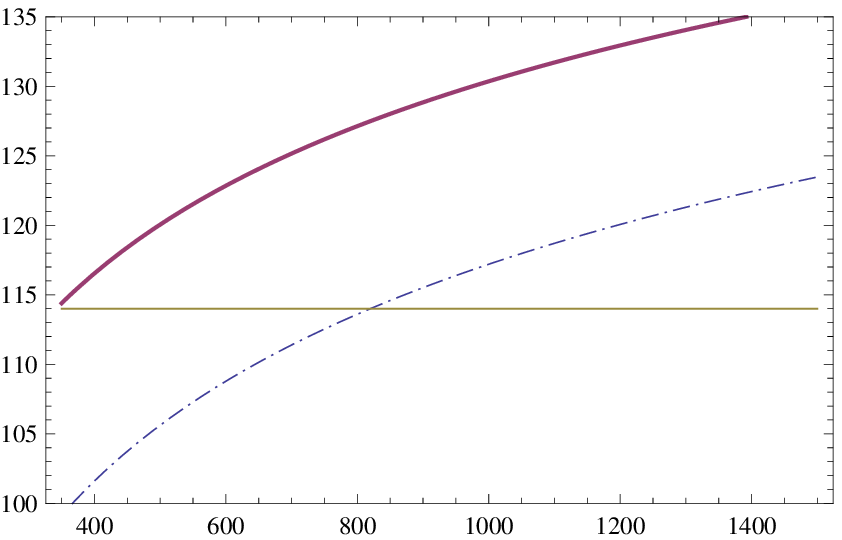}\\
~\hspace*{-0.1cm} {$M_S$}\hspace{6.8cm}{$M_S$}
\caption{{\it Left:} The dependence of the one--loop lightest Higgs boson mass
on $M_S$ for $\tan\beta=2$. {\it Right:} The one--loop mass of the lightest
CP--even Higgs state versus $M_S$ for $\tan\beta=3$. The solid and dashed--dotted
lines correspond to $X_t=2\,M_S$ and $X_t= M_S$ respectively. The horizontal line
represents the current LEP limit.}
\label{figure-1}
\end{figure}

Recent studies indicate that in the MSSM the scenarios with the light Higgs
pseudoscalar ($m_A\sim 100\,\mbox{GeV}$) are almost ruled out by LEP. Since
$m_A$ tend to be large the SM--like Higgs boson must be relatively heavy.
At the same time the lightest Higgs boson mass in the MSSM does not exceed
$M_Z\simeq 91\,\mbox{GeV}$ at the tree--level. As a consequence in order to
satisfy LEP constraints large contribution of loop corrections to
the mass of the lightest CP--even Higgs state is required. When SUSY breaking
scale $M_S$ is considerably larger than $M_Z$ and $m_Q^2\simeq m_U^2\simeq M_S^2$
the contribution of the one--loop corrections to $m_{h_1}^2$ in the leading
approximation can be written as
\begin{equation}
\Delta^{(1)}_{11}\simeq\frac{3 M_t^4 }{2\pi^2 v^2}
\left[\frac{X_t^2}{M_S^2}\biggl(1-\frac{1}{12}\frac{X_t^2}{M_S^2}\biggr)+
\ln\biggl(\frac{M_S^2}{m_t^2}\biggr)\right]\,.
\label{23}
\end{equation}
The large values of $\Delta^{(1)}_{11}\sim M_{Z}^2$ can be obtained only
if $M_S\gg m_t$ and the ratio $|X_t/M_S|$ is also large. The contribution of
the one--loop corrections (\ref{23}) attains its maximal value for $X_t^2=6\,M_S^2$.
This is the so--called maximal mixing scenario. In Figs.~1a and 1b the dependence of
the one--loop lightest Higgs boson mass on the SUSY breaking scale $M_S$ for $\tan\beta=2$
and $\tan\beta=3$ is examined. Two different cases $X_t=M_S$ and $X_t=2\,M_S$ are
considered. From Figs.~1a and 1b one can see that in order to satisfy LEP constraints
$M_S$ should be larger than $400-600\,\mbox{GeV}$. Leading two--loop corrections
reduce the SM--like Higgs mass even further. As a result larger values of the
SUSY breaking scale are required to overcome LEP limit.

Large values of the soft scalar masses of the superpartners of the top quark
($m_Q^2,\,m_U^2 \gg M_Z^2$) tend to induce large mass parameter $m_2^2$ at low
energies due to the RG flow. This leads to the fine tuning because $m_1^2$
and $m_2^2$ determine the EW scale (see Eqs.~(\ref{12})).
Generically the fine tuning which is required to overcome LEP constraints
in the MSSM is of the order of 1\% (little hierarchy problem). This fine
tuning should be compared with the fine tuning in other theories which are
used in particle physics. In particular, it is well known that QCD is
a highly fine--tuned theory. Indeed, QCD Lagrangian should
contain ``$\theta$-term''
\begin{equation}
{\cal L}_{\theta} = \theta_{\rm eff}
\frac{\alpha_s}{8 \pi} F^{\mu \nu \, a} \tilde{F}_{\mu \nu}^a,
\qquad\qquad \theta_{\rm eff} = \theta + {\rm arg \; det}\; M_q\,,
\label{24}
\end{equation}
where $F^{\mu \nu \, a}$ is the gluon field strength and
$\tilde{F}_{\mu \nu}^a \equiv\frac{1}{2}\epsilon_{\mu\nu\rho\sigma}F^{\rho\sigma\, a}$
is its dual. This term is not forbidden by the gauge invariance. On the other
hand the parameter $\theta$ must be extremely small, i.e.
$|\theta_{\rm eff}| \lesssim 10^{-9}$ (strong CP problem). Otherwise it results
in too large value of the neutron electric dipole moment. Eqs.~(\ref{24})
demonstrate that so small value of $\theta_{\rm eff}$ implies enormous
fine tuning which is much higher than in the MSSM.

The little hierarchy problem can be solved within SUSY models that allow to get
relatively large mass of the SM--like Higgs boson ($m_{h_1}\gtrsim 100-110\,\mbox{GeV}$)
at the tree-level. Alternatively, one can try to avoid stringent LEP constraints
by allowing exotic decays of the lightest Higgs particle. If usual branching ratios
of the lightest Higgs state are dramatically reduced then the lower LEP bound
on the SM--like Higgs mass may become inapplicable. In this case the lightest
Higgs boson can be still relatively light so that large contribution of
loop corrections is not required. Both possibilities mentioned above imply
the presence of new particles and interactions. These new particles and
interactions can be also used to solve the so-called $\mu$ problem.
This problem arises when MSSM gets incorporated into supergravity and/or
GUT models. Within these models the parameter $\mu$ is expected to be
either zero or of the order of Planck/GUT scale. At the same time in order
to provide the correct pattern of the EWSB $\mu$ is required to be of the
order of the EW scale.

\section{Higgs sector of the NMSSM}
In the simplest extension of the MSSM, the Next--to--Minimal Supersymmetric Standard
Model (NMSSM), the superpotential is invariant with respect to the discrete
transformations $\phi'_{\alpha}\to e^{2\pi i/3}\phi_{\alpha}$
of the $Z_3$ group (for recent review see \cite{Ellwanger:2009dp}). The term $\mu (H_1 H_2)$
does not meet this requirement. Therefore it is replaced in the superpotential by
\begin{equation}
W_{H}=\lambda S (H_1 H_2)+\frac{1}{3}\kappa S^3\,,
\label{25}
\end{equation}
where $S$ is an additional superfield which is a singlet with respect to $SU(2)_W$
and $U(1)_Y$ gauge transformations. A spontaneous breakdown of the EW symmetry
leads to the emergence of the VEV of extra singlet field $<S>=s/\sqrt{2}$ and an
effective $\mu$ parameter is generated ($\mu=\lambda s/\sqrt{2}$).

The potential energy of the Higgs field interaction can be written as a sum
\begin{equation}
V=V_F+V_D+V_{soft}+\Delta V\, ,
\label{26}
\end{equation}
\begin{equation}
V_F=\lambda^2|S|^2(|H_1|^2+|H_2|^2)+\lambda^2|(H_1 H_2)|^2+
\lambda\kappa\left[S^{*2}(H_1 H_2)+h.c.\right]+\kappa^2|S|^4\, ,
\label{27}
\end{equation}
\begin{equation}
V_D=\frac{g_2^2}{8}\left(H_1^+\sigma_a H_1+H_2^+\sigma_a
H_2\right)^2+\frac{{g'}^2}{8}\left(|H_1|^2-|H_2|^2\right)^2\, ,
\label{28}
\end{equation}
\begin{equation}
V_{soft}=m_1^2|H_1|^2+m_2^2|H_2|^2+m_S^2|S|^2
+\left[\lambda A_{\lambda}S(H_1 H_2)+\frac{\kappa}{3}A_{\kappa}S^3+h.c.\right]\, .
\label{29}
\end{equation}
At the tree level the Higgs potential (\ref{26}) is described by the sum of the first three terms.
$V_F$ and $V_D$ are the $F$ and $D$ terms. Their structure is fixed by the superpotential (\ref{25})
and the EW gauge interactions in the common manner. The soft SUSY breaking terms are collected in
$V_{soft}$. The set of soft SUSY breaking parameters involves soft masses $m_1^2,\, m_2^2,\, m_S^2$
and trilinear couplings $A_{\kappa},\, A_{\lambda}$. The last term in Eq.~(\ref{26}), $\Delta V$,
corresponds to the contribution of loop corrections. In the leading one--loop approximation
$\Delta V$ in the NMSSM is given by Eqs.~(\ref{14}) in which $\mu$ has to be replaced
by $\lambda s/\sqrt{2}$. Further we assume that $\lambda$, $\kappa$ and all soft SUSY breaking
parameters are real so that CP is conserved.

At the physical vacuum of the Higgs potential
\begin{equation}
<H_1>=\frac{1}{\sqrt{2}}\left(
\begin{array}{c}
v_1\\ 0
\end{array}
\right) , \qquad
<H_2>=\frac{1}{\sqrt{2}}\left(
\begin{array}{c}
0\\ v_2
\end{array}
\right)\,,\qquad <S>=\frac{s}{\sqrt{2}}\,.
\label{30}
\end{equation}
The equations for the extrema of the full Higgs boson effective potential in the directions
(\ref{30}) in the field space are given by
\begin{equation}
\frac{\partial V}{\partial s}=\left(m_S^2+\frac{\lambda^2}{2}(v_1^2+v_2^2)-\lambda\kappa v_1 v_2
\right)s-\frac{\lambda A_{\lambda}}{\sqrt{2}}v_1v_2+\frac{\kappa A_{\kappa}}{\sqrt{2}}s^2+
\kappa^2 s^3+\frac{\partial \Delta V}{\partial s}=0\, ,
\label{31}
\end{equation}
\begin{equation}
\frac{\partial V}{\partial v_1}=\left(m_1^2+\frac{\lambda^2}{2}(v_2^2+s^2)+
\frac{\bar{g}^2}{8}(v_1^2-v_2^2)\right)v_1-\left(\frac{\lambda\kappa}{2}s^2+
\frac{\lambda A_{\lambda}}{\sqrt{2}}s\right)v_2+\frac{\partial \Delta V}{\partial v_1}=0\, ,
\label{32}
\end{equation}
\begin{equation}
\frac{\partial V}{\partial v_2}=\left(m_2^2+\frac{\lambda^2}{2}
(v_1^2+s^2)+\frac{\bar{g}^2}{8}(v_2^2-v_1^2)\right)v_2-\left(\frac{\lambda\kappa}{2}s^2+
\frac{\lambda A_{\lambda}}{\sqrt{2}}s\right)v_1+\frac{\partial \Delta V}{\partial v_2}=0\, .
\label{33}
\end{equation}
As in the MSSM upon the breakdown of the EW symmetry three goldstone modes ($G^{\pm}$ and $G^{0}$)
emerge, and are absorbed by the $W^{\pm}$ and $Z$ bosons. In the field space basis rotated by an
angle $\beta$ with respect to the initial direction, i.e.
\begin{equation}
\begin{array}{ll}
H_1^-=G^- \cos \beta + H^- \sin \beta\, , \qquad          & H_2^+=H^+ \cos \beta - G^+ \sin \beta\, , \\
Im\, H_1^0 = (P \sin \beta + G^0 \cos \beta)/\sqrt{2}\, ,\qquad & Re \, H_1^0= (h \cos\beta-
H \sin\beta+v_1)/\sqrt{2}\,, \\
Im\, H_2^0 = (P \cos \beta - G^0 \sin \beta)/\sqrt{2} \, ,\qquad & Re \, H_2^0= (h \sin\beta+
H \cos\beta+v_2)/\sqrt{2}\,, \\
Im\, S = P_S/\sqrt{2}\,,     & Re \, S = (s+N)/\sqrt{2}\, ,
\end{array}
\label{34}
\end{equation}
these unphysical degrees of freedom decouple and the mass terms in the Higgs boson potential
can be written as follows
\begin{equation}
V_{mass} = M_{H^{\pm}}^2  H^+ H^- +
\frac{1}{2} (P \,\, P_S)\, \tilde{M}^2
\left(
\begin{array}{c}
P \\
P_S
\end{array}
\right)+
\frac{1}{2} (H \,\, h \,\, S)\, M^2
\left(
\begin{array}{c}
H \\
h \\
S
\end{array} \right)\, .
\label{35}
\end{equation}

From the conditions for the extrema (\ref{31})--(\ref{33}) one can express $m_S^2$, $m_1^2$,
$m_2^2$ via other fundamental parameters, $\tan\beta$ and $s$. Substituting the obtained
relations for the soft masses in the $2\times 2$ CP-odd mass matrix $\tilde{M}^2_{ij}$ we get:
$$
\tilde{M}_{11}^2=m_A^2=\frac{4\mu^2}{\sin^2 2\beta}\left(x-\frac{\kappa}{2\lambda}\sin2\beta\right)
+\tilde{\Delta}_{11}\, ,\qquad
\tilde{M}_{22}^2=\frac{\lambda^2 v^2}{2}x+\frac{\lambda\kappa}{2}v^2\sin2\beta-
3\frac{\kappa}{\lambda}A_{\kappa}\mu+\tilde{\Delta}_{22}\, ,
$$
\begin{equation}
\tilde{M}_{12}^2=\tilde{M}_{21}^2=\sqrt{2}\lambda v \mu\left(\frac{x}{\sin 2\beta}
-2\frac{\kappa}{\lambda}\right)+\tilde{\Delta}_{12}\, ,\\
\label{36}
\end{equation}
where $x=\displaystyle\frac{1}{2\mu}\left(A_{\lambda}+2\frac{\kappa}{\lambda}\mu\right)\sin2\beta$,
$\mu=\displaystyle\frac{\lambda s}{\sqrt{2}}$ and $\tilde{\Delta}_{ij}$
are contributions of the loop corrections to the mass matrix elements.
The mass matrix (\ref{36}) can be easily diagonalized.
The corresponding eigenvalues are given by
\begin{equation}
m^2_{A_2, A_1}=\frac{1}{2}\left(\tilde{M}^2_{11}+\tilde{M}^2_{22}\pm
\sqrt{(\tilde{M}^2_{11}-\tilde{M}^2_{22})^2+4\tilde{M}^4_{12}}
\right)~,
%\qquad  \tan \theta_A=\frac{\tilde{M}^2_{12}}{\tilde{M}^2_{11}-m^2_{A_1}}~.
\label{38}
\end{equation}

Because the charged components of the Higgs doublets are not mixed with the neutral Higgs
states the charged Higgs fields $H^{\pm}$ are already physical mass eigenstates with
\begin{equation}
M_{H^{\pm}}^2=m_A^2-\frac{\lambda^2 v^2}{2}+M_W^2+\Delta_{\pm}.
\label{39}
\end{equation}
Here $\Delta_{\pm}$ includes loop corrections to the charged Higgs mass.

In the rotated basis  $H\, ,h\,, N$ the mass matrix of the CP--even Higgs sector
takes the form \cite{Kovalenko:1998dc},\cite{Nevzorov:2001um}:
\begin{equation}
M^2=\left(
\begin{array}{ccc}
M_{11}^2 & M_{12}^2 & M_{13}^2\\[2mm]
M_{21}^2 & M_{22}^2 & M_{23}^2\\[2mm]
M_{31}^2 & M_{32}^2 & M_{33}^2
\end{array}
\right)=
\left(
\begin{array}{ccc}
\displaystyle\frac{1}{v^2}\frac{\partial^2 V}{\partial^2\beta}&
\displaystyle\frac{1}{v}\frac{\partial^2 V}{\partial v \partial\beta}&
\displaystyle\frac{1}{v}\frac{\partial^2 V}{\partial s \partial\beta}\\[2mm]
\displaystyle\frac{1}{v}\frac{\partial^2 V}{\partial v \partial\beta}&
\displaystyle\frac{\partial^2 V}{\partial v^2}&
\displaystyle\frac{\partial^2 V}{\partial v \partial s}\\[2mm]
\displaystyle\frac{1}{v}\frac{\partial^2 V}{\partial s \partial\beta}&
\displaystyle\frac{\partial^2 V}{\partial v \partial s}&
\displaystyle\frac{\partial^2 V}{\partial^2 s}
\end{array}
\right)~,
\label{40}
\end{equation}
\begin{equation}
\begin{array}{rcl}
M_{11}^2&=&\displaystyle m_A^2+\left(\frac{\bar{g}^2}{4}-\frac{\lambda^2}{2}\right)v^2
\sin^2 2\beta+\Delta_{11}\, ,\\
M_{22}^2&=&\displaystyle M_Z^2\cos^2 2\beta+\frac{\lambda^2}{2}v^2\sin^2 2\beta+
\Delta_{22}\, ,\\
M_{33}^2&=&\displaystyle 4\frac{\kappa^2}{\lambda^2}\mu^2+\frac{\kappa}{\lambda}A_{\kappa}\mu+
\frac{\lambda^2 v^2}{2}x-\frac{\kappa\lambda}{2}v^2\sin2\beta+\Delta_{33}\, ,\\
M_{12}^2&=&M_{21}^2=\displaystyle \left(\frac{\lambda^2}{4}-\frac{\bar{g}^2}{8}\right)v^2
\sin 4\beta+\Delta_{12}\, ,\\
M_{13}^2&=&M_{31}^2=-\sqrt{2}\lambda v \mu x\, \mbox{ctg} 2\beta+\Delta_{13}\, ,\\
M_{23}^2&=&M_{32}^2=\sqrt{2}\lambda v \mu (1-x)+\Delta_{23}\, ,
\end{array}
\label{41}
\end{equation}
where $\Delta_{ij}$ can be calculated by differentiating $\Delta V$.

At least one Higgs state in the CP--even sector is always light. Since the minimal eigenvalue of
a Hermitian matrix does not exceed its smallest diagonal element the lightest CP--even Higgs boson squared
mass $m_{h_1}^2$ remains smaller than $M_{22}^2\sim M_Z^2$ even when the supersymmetry breaking
scale is much larger than the EW scale\footnote{The same theorem may lead to the upper bound on the
mass of the lightest neutralino \cite{Hesselbach:2007te}.}. The upper bound on the lightest Higgs mass 
in the NMSSM differs from the theoretical bound in the MSSM only for moderate values of $\tan\beta$. 
In the leading two--loop approximation the lightest Higgs boson mass in the NMSSM does not exceed 
$135\,\mbox{GeV}$.
%As in the MSSM the trilinear part of the Lagrangian, which determines the interactions of the neutral
%Higgs states with the $Z$--boson, is given by Eq.~(\ref{20}). The couplings of the Higgs scalars to
%the Z pair ($R_{ZZi}$, $i=1,2,3$) and to the Higgs pseudoscalars and Z boson ($R_{ZA_1i}$ and $R_{ZA_2i}$)
%appear because of the mixing of $h, H$ and $P$ with other components of the CP--odd and CP--even Higgs
%sectors. $H,\, h$ and $N$ are related to the physical CP-even Higgs eigenstates by virtue of unitary
%transformation:
%\begin{equation}
%\left(
%\begin{array}{c}
%H\\ h\\ N
%\end{array}
%\right)=
%U^{+}
%\left(
%\begin{array}{c}
%h_3\\ h_2\\ h_1
%\end{array}
%\right)\, .
%\label{42}
%\end{equation}
%Combining Lagrangian (\ref{20}) and relations (\ref{37}) and (\ref{42}) the normalized $R$--couplings may be
%written in terms of the mixing  matrices
%\begin{equation}
%R_{ZZi}=U^+_{hi}~,~~~R_{ZA_{1}i}=-U^+_{Hi}\sin\theta_A~,~~~
%R_{ZA_2i}=U^+_{Hi}\cos\theta_A~.
%\label{43}
%\end{equation}
As follows from the explicit form of the mass matrices (\ref{36}) and (\ref{41}) at the tree-level,
the spectrum of the Higgs bosons and their couplings depend on the six parameters:
$\lambda, \kappa, \mu, \tan\beta, A_{\kappa}$ and $m_A$ (or $x$). 

First let us consider the MSSM limit of the NMSSM.
Because the strength of the interaction of the extra SM singlet superfield $S$ with $H_1$ and $H_2$
is determined by the size of the coupling $\lambda$ in the superpotential (\ref{25}) the MSSM expressions
for the Higgs masses and couplings are reproduced when $\lambda$ tends to be zero. On the other hand
the equations (\ref{32})--(\ref{33}) imply that $s$ should grow with decreasing $\lambda$ as $M_Z/\lambda$
to ensure the correct breakdown of the EW symmetry. In the limit $\lambda\to 0$ all terms, which are
proportional to $\lambda v_i$, in the minimization conditions (\ref{31}) can be neglected and the
corresponding equation takes the form:
\begin{equation}
s\left(m_S^2+\frac{\kappa A_{\kappa}}{\sqrt{2}}s
+\kappa^2 s^2\right)\simeq 0
\label{44}
\end{equation}
The Eq.~(\ref{44}) has always at least one solution $s_0=0$. In addition two non-trivial roots arise
if $A_{\kappa}^2> 8 m_S^2$. They are given by
\begin{equation}
s_{1,2}\simeq\frac{-A_{\kappa}\pm\sqrt{A_{\kappa}^2-8 m_S^2}}{2\sqrt{2}\kappa}\,\,.
\label{45}
\end{equation}
When $m_S^2>0$ the root $s_0=0$ corresponds to the local minimum of the Higgs potential (\ref{26})--(\ref{29})
that does not lead to the acceptable solution of the $\mu$--problem. The second non-trivial vacuum, that appears if
$A_{\kappa}^2> 8 m_S^2$, remains unstable for $A_{\kappa}^2< 9 m_S^2$. Larger absolute values of $A_{\kappa}$
$(A_{\kappa}^2> 9 m_S^2)$ stabilizes the second minimum which is attained at $s=s_1 (s_2)$ for negative
(positive) $A_{\kappa}$. From Eq.~(\ref{45}) it becomes clear that the increasing of $s$ can be achieved either
by decreasing $\kappa$ or by raising $m_S^2$ and $A_{\kappa}$. Since there is no natural reason why $m_S^2$ and
$A_{\kappa}$ should be very large while all other soft SUSY breaking terms are left in the TeV range, the values
of $\lambda$ and $\kappa$ are obliged to go to zero simultaneously so that their ratio remains unchanged.

Since in the MSSM limit of the NMSSM mixing between singlet states and neutral components of the Higgs doublets
is small the mass matrices (\ref{36}) and (\ref{40})--(\ref{41}) can be diagonalised using the perturbation
theory \cite{Kovalenko:1998dc},\cite{Nevzorov:2001um},\cite{Nevzorov:2000uv}--\cite{Nevzorov:2004ge}.
At the tree--level the masses of two Higgs pseudoscalars are given by
\begin{equation}
m_{A_2}^2\simeq m_A^2=\frac{4\mu^2}{\sin^2 2\beta}\left(x-\frac{\kappa}{2\lambda}\sin2\beta\right)\,, \qquad\qquad\qquad m_{A_1}^2\simeq-3\frac{\kappa}{\lambda}A_{\kappa}\mu\,.
\label{46}
\end{equation}
The masses of two CP--even Higgs bosons are the same as in the MSSM (see Eq.~(\ref{19})) while the mass of the extra
CP--even Higgs state, which is predominantly a SM singlet field, is set by $\frac{\kappa}{\lambda}\mu$
\begin{equation}
m_{h_3}^2\approx 4\frac{\kappa^2}{\lambda^2}\mu^2+\frac{\kappa}{\lambda}A_{\kappa}\mu
+\frac{\lambda^2v^2}{2}x\sin^22\beta-\frac{2\lambda^2v^2\mu^2(1-x)^2}{M_Z^2\cos^22\beta}\, .
\label{47}
\end{equation}
The parameter $A_{\kappa}$ occurs in the masses of extra scalar $m_{h_3}$ and pseudoscalar $m_{A_1}$ with opposite
sign and is therefore responsible for their splitting. To ensure that the physical vacuum is a global minimum of the
Higgs potential (\ref{26})--(\ref{29}) and the masses-squared of all Higgs states are positive in this
vacuum the parameter $A_{\kappa}$ must satisfy the following constraints
\begin{equation}
-3\left(\frac{\kappa}{\lambda}\mu\right)^2 \lesssim A_{\kappa}\left(\frac{\kappa}{\lambda}\mu\right)\lesssim 0\, .
\label{48}
\end{equation}
The experimental constraints on the SUSY parameters obtained in the MSSM remain valid in the NMSSM with small
$\lambda$ and $\kappa$. For example, non--observation of any neutral Higgs particle and chargino at the LEP II
imply that $\tan\beta\gtrsim 2.5$ and $|\mu|\gtrsim 90-100\,\mbox{GeV}$.

Decreasing $\kappa$ reduces the masses of extra scalar and pseudoscalar states so that for $\kappa\ll\lambda$
they can be the lightest particles in the Higgs boson spectrum. In the limit $\kappa\to 0$ the mass of the
lightest pseudoscalar state vanishes. In the considered limit the Lagrangian of the NMSSM is invariant under
transformations of $SU(2)\times [U(1)]^2$ global symmetry. Extra $U(1)$ global symmetry gets spontaneously
broken by the VEV of the singlet field $S$, giving rise to a massless Goldstone boson, the Peccei--Quinn (PQ)
axion. In the PQ--symmetric NMSSM astrophysical observations exclude any choice of the parameters unless
one allows $s$ to be enormously large ($>10^{9}-10^{11}\,\mbox{GeV}$). These huge vacuum expectation values of
the singlet field can be consistent with the EWSB only if $\lambda\sim 10^{-6}-10^{-9}$
\cite{Miller:2003hm}--\cite{Miller:2005qua}. Therefore here we restrict our consideration to small but non--zero 
values of $\kappa\lesssim \lambda^2$ when the PQ-symmetry is only slightly broken.

As evident from Eq.~(\ref{47}) at small values of $\kappa$ the mass--squared of the lightest Higgs scalar
tends to be negative if $|\mu|$ is large and/or the auxiliary variable $x$ differs too much from unity.
Due to the vacuum stability requirement, which implies the positivity of the mass--squared of all Higgs
particles, $x$ has to be localized near unity, i.e.
\begin{equation}
1-\left|\frac{\sqrt{2}\kappa M_Z}{\lambda^2 v}\right|<x<1+\left|\frac{\sqrt{2}\kappa M_Z}{\lambda^2 v}\right|\,.
\label{49}
\end{equation}
This leads to the hierarchical structure of the Higgs spectrum. Indeed, combining LEP limits on $\tan\beta$ and
$\mu$ one gets that $m_A^2\gtrsim 9M_Z^2\, x$. Because of this the heaviest CP--odd, CP--even and charged Higgs bosons
are almost degenerate with masses around $m_A\simeq \mu\tan\beta$ while the SM--like Higgs state has a mass of
the order of $M_Z$.

\begin{figure}
\centering
\includegraphics[width=.6\textwidth]{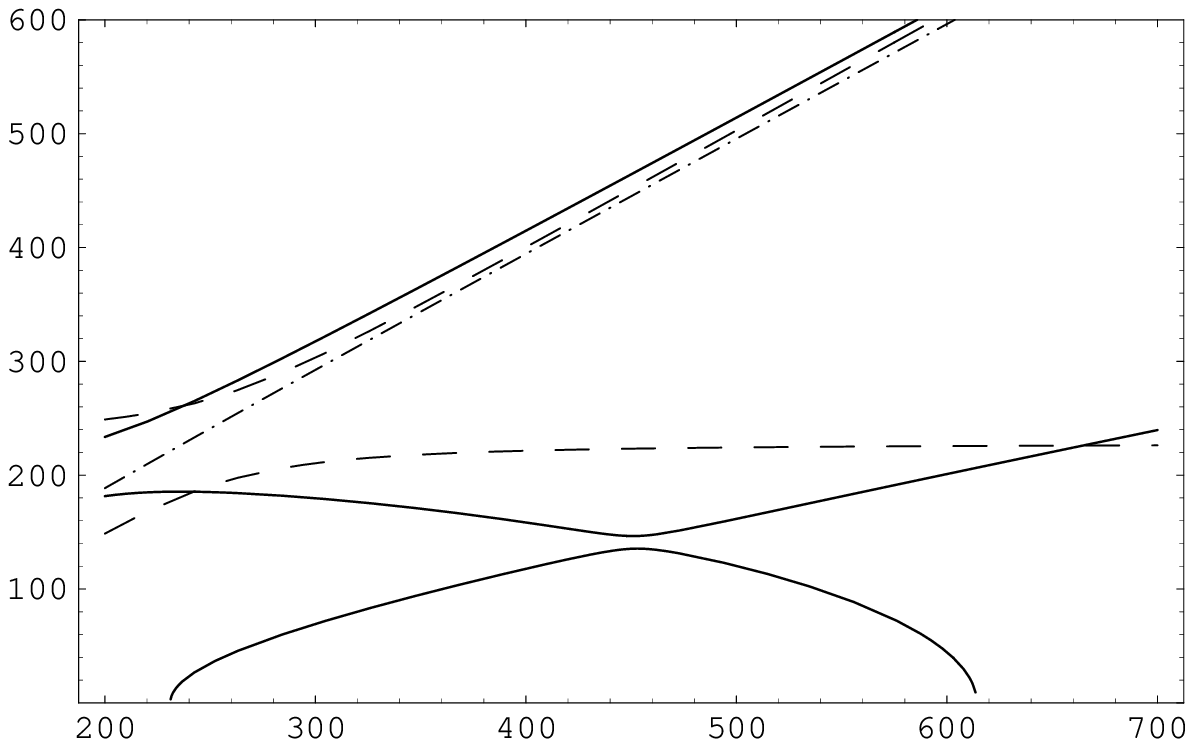}\\
{$m_A (\mbox{GeV})$}
\caption{One--loop masses of the CP--even Higgs bosons
versus $m_A$ for $\lambda=0.6$, $\kappa=0.36$, $\tan\beta=3$,
$\mu=150\,\mbox{GeV}$, $A_{\kappa}=135\,\mbox{GeV}$,
$m_Q^2=m_U^2=M_S^2$, $X_t=\sqrt{6} M_S$ and $M_S=700\,\mbox{GeV}$.
Solid, dashed and dashed--dotted lines correspond to the masses of
the CP--even, CP--odd and charged Higgs bosons respectively.}
\label{figure-2}
\end{figure}

The main features of the NMSSM Higgs spectrum discussed above are retained when the couplings
$\lambda$ and $\kappa$ increase. For the appreciable values of $\kappa$ and $\lambda$ the
slight breaking of the PQ--symmetry can be caused by
%small ratio $\kappa/\lambda$ may arise due to
the RG flow of these couplings from the GUT scale $M_X$ to $M_Z$. In the infrared region the
solutions of the NMSSM RG equations are focused near the intersection of the Hill-type effective
surface and invariant line \cite{Nevzorov:2001vj}--\cite{Nevzorov:2002ub}. As a result
at the EW scale $\kappa/\lambda$ tend to be less than unity even when $\kappa(M_X) > \lambda(M_X)$
initially. In Fig.~2 the dependence of the masses and couplings of the Higgs bosons on $m_A$ is
examined. As a representative example we fix the Yukawa couplings so that
$\lambda(M_X)=\kappa(M_X)=2 h_t(M_X)=1.6$, that corresponds to $\tan\beta\gtrsim 3$, $\lambda(M_t)=0.6$
and $\kappa(M_t)=0.36$. In order to obtain a realistic spectrum, we include the leading one--loop
corrections from the top and stop loops. From Fig.~2 it becomes clear that the requirement of
stability of the physical vacuum limits the range of variations of $m_A$ from below and above
maintaining the mass hierarchy in the Higgs spectrum. Relying on this mass hierarchy the
approximate solutions for the Higgs masses and couplings can be
obtained \cite{Miller:2003ay},\cite{Nevzorov:2004ge}. The numerical results in
Fig.~2 reveal that the masses of the heaviest CP--even, CP--odd and charged Higgs states are
approximately degenerate while the other three neutral states are considerably lighter.
The hierarchical structure of the Higgs spectrum ensures that the heaviest CP--even and CP--odd
Higgs bosons are predominantly composed of $H$ and $P$. As before the lightest Higgs scalar and
pseudoscalar are singlet dominated, making their observation quite problematic.  The second
lightest CP--even Higgs boson has a mass around $130\,\mbox{GeV}$, mimicking the lightest Higgs
scalar in the MSSM. Observing two light scalars and one pseudoscalar Higgs particles but no charged
Higgs boson at future colliders would yield an opportunity to differentiate the NMSSM with a slightly
broken PQ--symmetry from the MSSM even if the heavy Higgs states are inaccessible.

The presence of light singlet scalar and pseudoscalar permits to weaken the LEP lower bound
on the lightest Higgs boson mass. These states have reduced couplings to Z--boson that could
allow them to escape the detection at LEP. On the other hand singlet scalar can mix with
the SM-like superposition $h$ of the neutral components of Higgs doublets resulting in the
reduction of the couplings of the second lightest Higgs scalar to $Z$--boson. This relaxes
LEP constraints so that the SM-like Higgs state does not need to be considerably heavier
than $100\,\mbox{GeV}$. Therefore large contribution of loop corrections to the mass of
the SM-like Higgs boson is not required. Another possibility to overcome the little hierarchy
problem is to allow the SM--like Higgs state to decay predominantly into two light
singlet pseudoscalars $A_1$ (for recent review see \cite{Chang:2008cw}). This can be achieved
because the coupling of the SM--like Higgs boson to the $b$--quark is rather small. If this coupling
is substantially smaller than the coupling of the SM--like Higgs state to $A_1$ then the decay mode
$h\to A_1 + A_1$ dominates. The singlet pseudoscalar can sequentially decay into either
$b\bar{b}$ or $\tau\bar{\tau}$ leading to four fermion decays of the SM--like Higgs boson.
In this case, again, the corresponding Higgs eigenstate might be relatively light that
permits to avoid little hierarchy problem.

\begin{figure}
\centering
~\hspace*{-6.1cm}{$m_{h_1}$}\hspace{6.8cm}{$m_{h_1}$}\\
\includegraphics[width=0.43\textwidth]{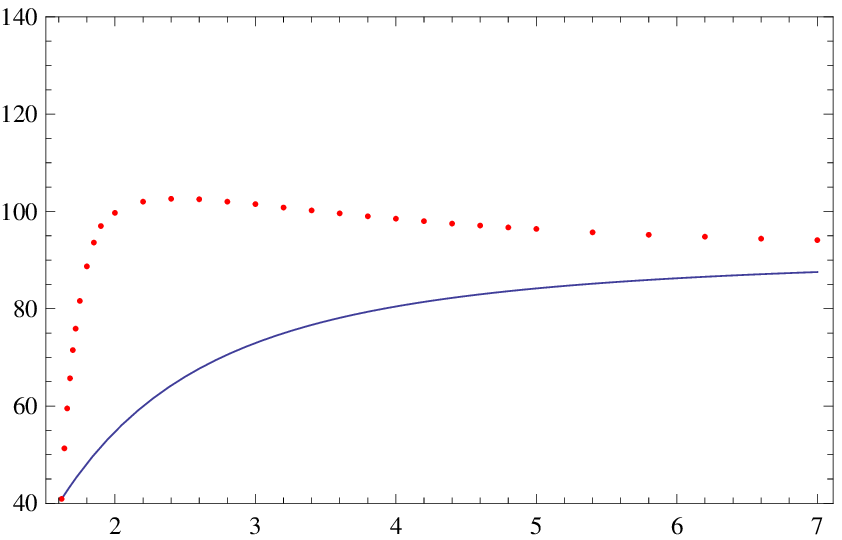}\qquad
\includegraphics[width=0.43\textwidth]{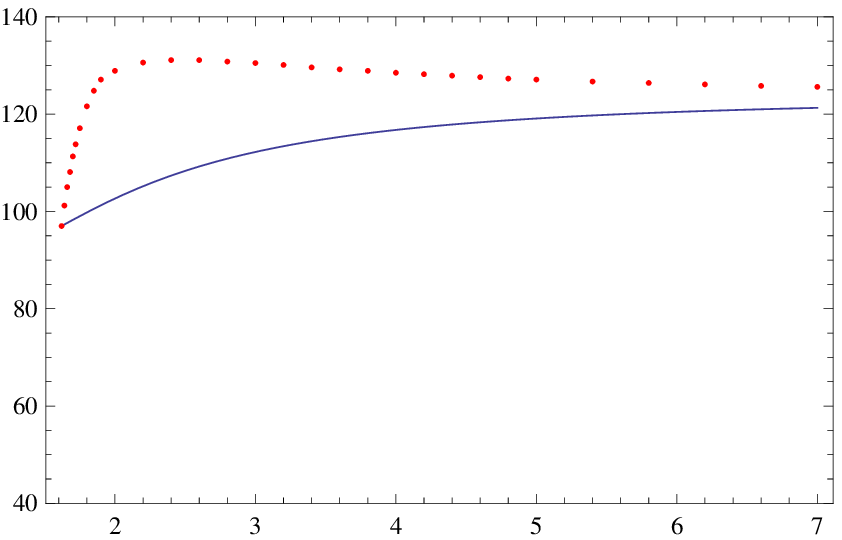}\\
~\hspace*{-0.1cm} {$\tan\beta$}\hspace{6.8cm}{$\tan\beta$}
\caption{
{\it Left:} Tree--level upper bound on the lightest Higgs boson mass
in the MSSM and NMSSM as a function of $\tan\beta$.
{\it Right:} The dependence of the two--loop upper bound on the lightest
Higgs boson mass on $\tan\beta$ for $m_t(M_t)=165\,\mbox{GeV}$,
$m_Q^2=m_U^2=M_S^2$, $X_t=\sqrt{6} M_S$ and $M_S=700\,\mbox{GeV}$.
The solid and dotted lines represent the theoretical restrictions on
$m_{h_1}$ in the MSSM and NMSSM respectively.}
\label{figure-2}
\end{figure}

However even when the couplings of the lightest CP--even Higgs state are almost the same
as in the SM it is substantially easier to overcome LEP constraint on the mass of the
SM--like Higgs boson in the NMSSM than in the MSSM. Indeed, in the NMSSM the theoretical
upper bound on $m_{h_1}^2$, which is given by $M_{22}^2$ in Eq.~(\ref{41}), contains
an extra term $\frac{\lambda^2}{2}v^2\sin^2 2\beta$ which is not present in the MSSM.
Due to this term the maximum possible value of the mass of the lightest Higgs scalar
in the NMSSM can be considerably larger as compared with the MSSM 
at moderate values of $\tan\beta$. In our analysis we
require the validity of perturbation theory up to the scale $M_X$. This sets stringent
upper limit on $\lambda(M_t)$ at low energies for each particular choice of
$\tan\beta$. Using theoretical restrictions on $\lambda(M_t)$ one can compute the
the maximum possible value of $m_{h_1}^2$ for each given value of $\tan\beta$.
Fig.~3 shows the dependence of the upper bound on the lightest Higgs boson mass as a
function of $\tan\beta$ in the MSSM and NMSSM. From Fig.~3 one can see that at the
tree--level the lightest CP--even Higgs state in the NMSSM can be considerably heavier
than in the MSSM at moderate values of $\tan\beta$. As a consequence in the leading
two--loop approximation it is substantially easier to get $m_{h_1}\gtrsim 114.4\,\mbox{GeV}$
in the NMSSM than in the MSSM for $\tan\beta=2-4$.

\section{Higgs spectrum in the $E_6$ inspired SUSY models with extra $U(1)'$ factor}
Another solution to the $\mu$ problem arises within superstring inspired models
based on the $E_6$ gauge group. At high energies $E_6$ can be broken
$SU(3)_C\times SU(2)_W\times U(1)_Y\times U(1)'$. An extra $U(1)'$ that appears at
low energies is a linear superposition of $U(1)_{\chi}$ and $U(1)_{\psi}$:
\begin{equation}
U(1)'=U(1)_{\chi}\cos\theta+U(1)_{\psi}\sin\theta\,,
\label{50}
\end{equation}
where two anomaly--free $U(1)_{\psi}$ and $U(1)_{\chi}$ symmetries are defined by:
$E_6\to SO(10)\times U(1)_{\psi}$\,, $SO(10)\to SU(5)\times U(1)_{\chi}$.
If $\theta\ne 0$ or $\pi$ the extra $U(1)'$ gauge symmetry forbids an elementary $\mu$
term but allows interaction $\lambda S(H_1 H_2)$ in the superpotential. After
EWSB the scalar component of the SM singlet superfield $S$ acquires a non--zero VEV
breaking $U(1)'$ and an effective $\mu$ term of the required size is automatically
generated.

The Higgs sector of the considered models includes two Higgs doublets as well
as a SM--like singlet field $S$ that carries $U(1)'$ charge. The Higgs effective
potential can be written as
\begin{equation}
\begin{array}{rcl}
V&=&V_F+V_D+V_{soft}+\Delta V\, ,\\[1mm]
V_F&=&\lambda^2|S|^2(|H_d|^2+|H_u|^2)+\lambda^2|(H_d H_u)|^2\,,\\[1mm]
V_D&=&\frac{g_2^2}{8}\left(H_d^\dagger \sigma_a H_d+H_u^\dagger \sigma_a
H_u\right)^2+\frac{{g'}^2}{8}\left(|H_d|^2-|H_u|^2\right)^2\\[2mm]
&&+\frac{g^{'2}_1}{2}\left(\tilde{Q}_1|H_d|^2+\tilde{Q}_2|H_u|^2+\tilde{Q}_S|S|^2\right)^2\,,\\[1mm]
V_{soft}&=&m_{S}^2|S|^2+m_1^2|H_d|^2+m_2^2|H_u|^2+
\biggl[\lambda A_{\lambda}S(H_u H_d)+h.c.\biggr]\,,
\end{array}
\label{51}
\end{equation}
where $g'_1$ is $U(1)'$ gauge coupling and $\tilde{Q}_1$, $\tilde{Q}_2$ and
$\tilde{Q}_S$ are effective $U(1)'$ charges of $H_1$, $H_2$ and $S$ respectively.
In Eq.~(\ref{51}) $V_F$ and $V_D$ are the $F$ and $D$ terms, $V_{soft}$ contains
a set of soft SUSY breaking terms while $\Delta V$ represents the contribution of
loop corrections.

At the physical vacuum the Higgs fields acquire VEVs given by Eq.~(\ref{30})
thus breaking the $SU(2)_W\times U(1)_Y\times U(1)'$ symmetry to $U(1)_{em}$. 
As a result two CP--odd and two charged Goldstone
modes in the Higgs sector are absorbed by the $Z$, $Z'$ and $W^{\pm}$ gauge bosons
so that only six physical degrees of freedom are left. They form one CP--odd, three
CP--even and two charged states. The masses of the CP--odd and charged Higgs bosons
can be written as
\begin{equation}
m_A^2=\frac{2\lambda^2 s^2 x}{\sin^2 2\beta}+O(M_Z^2)\,,\qquad\qquad m^2_{H^{\pm}}=m_A^2+O(M_Z^2)\,,
\label{52}
\end{equation}
where $x=\frac{A_{\lambda}}{\sqrt{2}\lambda s}\sin 2\beta$\,. The masses of two heaviest
CP--even states are set by $M_{Z'}$ and $m_A$, i.e.
\begin{equation}
m^2_{h_3}=m_A^2+O(M_Z^2)\,,\qquad\qquad m^2_{h_2}=M_{Z'}^2+O(M_Z^2)\,,
\label{53}
\end{equation}
where $M_{Z'}\simeq g^{'}_1\tilde{Q}_S s$. The lightest CP--even Higgs boson has a mass which
is less than
\begin{equation}
m^2_{h_1}\lesssim \frac{\lambda^2}{2}v^2\sin^22\beta+M_{Z}^2\cos^22\beta+
g^{'2}_1 v^2\biggl(\tilde{Q}_1\cos^2\beta+\tilde{Q}_2\sin^2\beta\biggr)^2+\Delta\,.
\label{54}
\end{equation}
In Eq.~(\ref{54}) $\Delta$ represents the contribution of loop corrections. Since the mass of
the $Z'$ boson in the $E_6$ inspired models has to be heavier than $800-900\,\mbox{GeV}$
at least one CP--even Higgs state, which is singlet dominated, is always heavy.
If $m_A < M_{Z'}$ then we get MSSM--type Higgs spectrum. When $m_A > M_{Z'}$
the heaviest CP--even, CP--odd and charged states are almost degenerate
with masses around $m_A$. In this case the lightest Higgs state is predominantly the
SM-like superposition $h$ of the neutral components of Higgs doublets.

Recently the detailed analysis of the Higgs sector was performed within a particular
$E_6$ inspired SUSY model with an extra $U(1)_N$ gauge symmetry that corresponds to
$\theta=\arctan\sqrt{15}$ \cite{King:2005jy}-\cite{King:2005my}.
The extra $U(1)_N$ gauge symmetry is defined such that
right--handed neutrinos do not participate in the gauge interactions. Only in this
Exceptional Supersymmetric Standard Model (E$_6$SSM) right--handed may be superheavy,
shedding light on the origin of the mass hierarchy in the lepton sector and
providing a mechanism for the generation of the baryon asymmetry in the Universe via
leptogenesis \cite{King:2008qb}. To ensure anomaly cancellation the particle content
of the E$_6$SSM is extended to include three complete fundamental $27$ representations
of $E_6$. In addition to the complete $27_i$ multiplets the low energy particle spectrum
of the E$_6$SSM is supplemented by $SU(2)_W$ doublet $H'$ and anti-doublet $\overline{H}'$
states from extra $27'$ and $\overline{27'}$ to preserve gauge coupling unification.
The unification of gauge couplings in the considered model can be achieved for any
phenomenologically acceptable value of $\alpha_3(M_Z)$ consistent with the measured low
energy central value \cite{King:2007uj}. The Higgs spectrum within the E$_6$SSM was
studied in \cite{King:2005jy}-\cite{King:2005my}, \cite{King:2006vu}--\cite{King:2006rh}.
It was argued that even at the tree level the lightest Higgs boson
mass in this model can be larger than $120\,\mbox{GeV}$. Therefore nonobservation of the
Higgs boson at LEP does not cause any trouble for the E$_6$SSM, even at tree--level.
In the leading two--loop approximation the mass of the lightest CP--even Higgs boson
in the considered model does not exceed $150-155\,\mbox{GeV}$ \cite{King:2005jy}.
The presence of light exotic particles in the E$_6$SSM spectrum lead to the nonstandard
decays of the SM--like Higgs boson which were discussed in \cite{10}.

\end{document}